\documentclass[a4paper,10pt]{article}
\usepackage[top=0.5in,body={6.5in,10.1in}]{geometry}
\usepackage{amssymb,amsmath}
\usepackage{graphicx}
\usepackage{epsfig}
\usepackage{indentfirst}
\usepackage{natbib}
\begin{document}

\title{On the effect of turbulent anisotropy on pulsation stability of stars}
\author{
Chunguang Zhang\,$^1$, Licai Deng\,$^1$, Darun Xiong\,$^{2}$\\
{\it\small $^1$\,Key Laboratory of Optical Astronomy, National Astronomical Observatories,}\\
{\it\small Chinese Academy of Sciences, Beijing 100101, China}\\
{\it\small $^2$\,Purple Mountain Observatory, Chinese Academy of Sciences, Nanjing 210008, China}\\
}
\date{}
\maketitle

\begin{abstract}
Within the framework of non-local time-dependent stellar convection theory, we study in detail the effect of turbulent anisotropy on stellar pulsation stability. The results show that anisotropy has no substantial influence on pulsation stability of g modes and low-order (radial order $n_\mathrm{r}<5$) p modes. The effect of turbulent anisotropy increases as the radial order increases. When turbulent anisotropy is neglected, most of high-order ($n_\mathrm{r}>5$) p modes of all low-temperature stars become unstable. Fortunately, within a wide range of the anisotropic parameter $c_3$, stellar pulsation stability is not sensitive to the specific value of $c_3$. Therefore it is safe to say that calibration errors of the convective parameter $c_3$ do not cause any uncertainty in the calculation of stellar pulsation stability.
\end{abstract}

\section{Introduction}
\label{sect:intro}
Radiation and convection are the two main mechanisms of energy transport in stellar interiors. In low-temperature late-type stars, convection, instead of radiation, becomes the dominant energy transport mechanism. Convection causes transport and exchange of energy, momentum, and material inside stars, and therefore has important influence on stellar structure, evolution and pulsation stability.

Non-locality and anisotropy are two most important properties of stellar convection. There has been much research and discussion regarding the effect of non-local convection \citep{XDC1998a, XDC1998b, XD2013}, while the effect of turbulent anisotropy on stellar pulsation stability is less studied. The purpose of this work is to provide an in-depth study and analysis about this problem. In Section \ref{sect:anisotropy} we give a brief introduction of our theoretical treatment of turbulent anisotropy. The dependence of stellar pulsation stability on the anisotropic parameter $c_3$ is discussed in Section \ref{sect:dependence} by means of numerical methods. The results are summarised in Section \ref{sect:conclusions}.

\section{The theoretical treatment of turbulent anisotropy}
\label{sect:anisotropy}
Convection is the internal instability of fluid medium induced by thermal instability. When the temperature gradient exceeds the adiabatic value in a gravitationally stratified fluid, rising and sinking fluid elements gain extra kinetic energy under buoyancy force, and convection occurs. The initial convective motion is along the direction of the gravitational field, i.e. the radial direction. However, a part of the kinetic energy of radial motion will be converted into that of horizontal motion as a result of the continuity of fluid motion and non-linear interaction among turbulence elements. The scientific study of turbulence began more than a century ago, but the nature of anisotropic turbulence has not been fully understood. There is no universally accepted anisotropic stellar convection theory. Different treatments of anisotropy of stellar convection have been adopted by different authors \citep{Unno1967, Gough1977, Canuto1993, LY2001}. We proposed a more general treatment of turbulent anisotropy \citep{DXC2006}, where the correlation of turbulent velocity fluctuation $\overline{w'^iw'^j}$ was decomposed into the isotropic component $g^{ij}x^2$ and the anisotropic component $\chi^{ij}$:
\begin{equation}
\overline{w'^iw'^j}=g^{ij}x^2+\chi^{ij}.
\label{eq1}
\end{equation}
From turbulence theory, we know that in the dynamic equations of turbulent velocity correlations, the correlation of pressure and velocity gradient tends to make turbulence isotropic \citep{Rotta1951}. Therefore we further assume that it has the form $-c_3\chi^{ij}/\tau_\mathrm{c}$, where $\tau_\mathrm{c}$ is the time scale of turbulent dissipation, and $c_3 \in (0,\infty)$ is an adjustable convective parameter. As $c_3$ increases, the pressure-velocity gradient correlation is more capable of restoring isotropy of turbulence, and so turbulence becomes more isotropic. The complete set of equations of stellar structure and pulsation under non-local and anisotropic time-dependent convection theory is described by \citet[][hereafter Paper I]{XDZ2015}. Asymptotic analysis shows that the ratio of mean squared radial velocity to mean squared horizontal velocity $\overline{u'^2_\mathrm{r}}/\overline{u'^2_\mathrm{h}} \approx (c_3+3)/2c_3$ in convectively unstable regions. This theoretical expectation from asymptotic analysis has been verified by numerical calculations, therefore $c_3$ is a convective parameter representing turbulent anisotropy. In the next section, we quantitatively study the effect of turbulent anisotropy on pulsation stability by calculating non-adiabatic oscillations.

\section{Dependence of pulsation stability on $\lowercase{c}_3$}
\label{sect:dependence}
\begin{figure}[htbp]
\centering
\includegraphics[width=\textwidth]{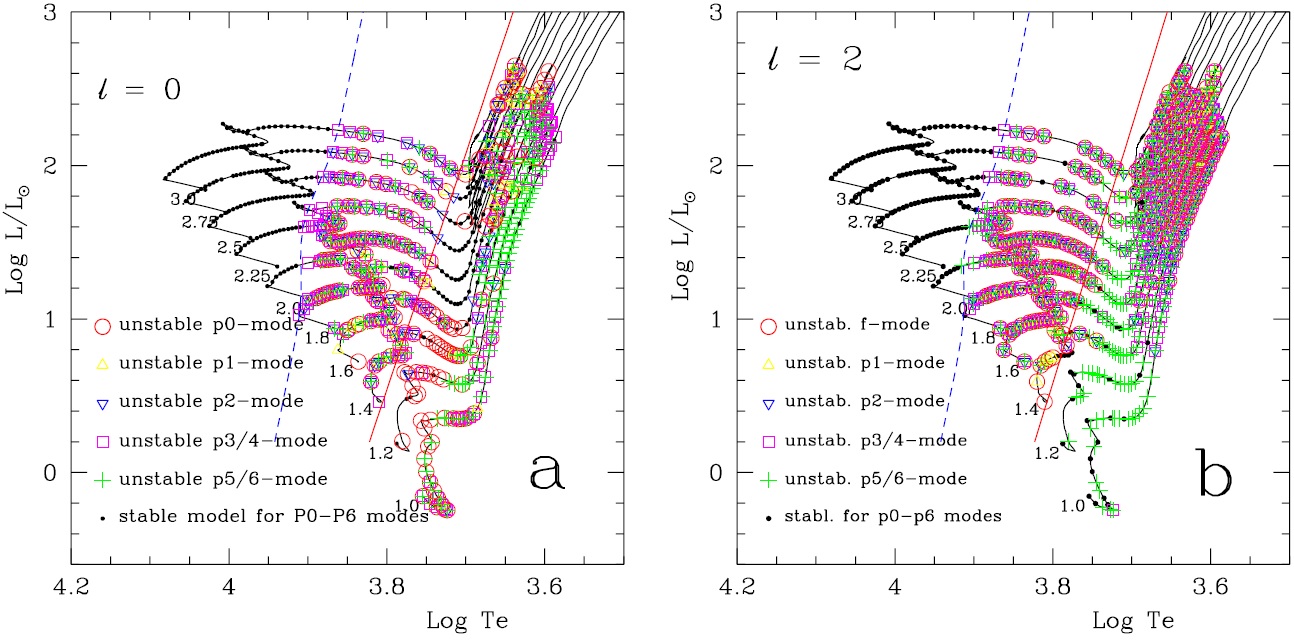}
\caption{Amplitude spectra of the four stars. The inserts are extended views of the low-frequencies.}
\label{fig1}
\end{figure}

We have briefly introduced our theoretical treatment of turbulent anisotropy in previous section. The degree of turbulent anisotropy can be calibrated by the convective parameter $c_3$. In Paper I, we derived a complete set of dynamic equations for the calculation of stellar structure and oscillations in non-local and anisotropic time-dependent convection theory. The equations return to isotropic convection approximation when the anisotropic component of the turbulent velocity correlation is set to zero, i.e. $\chi^{ij} = 0$, and the number of independent variables reduces from 8 to 7. Using Padova code \citep{Padova1993}, we calculated stellar evolutionary modes with $M=1.4$--$3.0M_\odot$, $X=0.70$, and $Z=0.02$. Following the procedure described in Paper I, we then calculated non-local envelope models in isotropic convection approximation as well as their radial and non-radial oscillations for g9--p29 modes. A modified version of MHD equation of state \citep{MHD1, MHD2, MHD3} and OPAL tabular opacity \citep{OPAL1992} complemented by low-temperature opacity \citep{AF1994} were adopted in the calculations. Figures \ref{fig1}a \& b show the distribution of stable (black dots) and unstable (colored open symbols) low-order p modes in the Hertzsprung-Russell (H-R) diagram for radial (the degree $l=0$) and $l=2$ non-radial oscillations, respectively. The dashed and solid lines are respectively the blue and red edges of the $\delta$ Scuti instability strip derived under anisotropic convection theory \citep{XDZW2016}. It is clear that there is no red edge of the instability strip when turbulent anisotropy is neglected in term of our theory.

Starting from the dynamic equation of stellar structure and oscillations using our non-local and anisotropic convection theory \citep{XDZ2015}:
\begin{equation} \frac{\mathrm{D}u^i}{\mathrm{D}t}+\frac{1}{\bar\rho}\nabla_k\left[g^{ik}\left(\bar P+\bar\rho\right)+\bar\rho\chi^{ik}\right]+g^{ik}\nabla_k\bar\phi=0,
\label{eq2}
\end{equation}
after linearization, we have
\begin{equation}
-\omega^2\delta r^i+\frac{1}{\bar\rho}\left\{\nabla_k\left[g^{ik}\left(\bar P+\bar\rho x^2\right)'+\left(\bar\rho \chi^{ik}\right)'\right]-\frac{\bar\rho'}{\bar\rho}\nabla_k\left[g^{ik}\left(\bar P+\bar\rho x^2\right)+\bar\rho\chi^{ik}\right]\right\}+g^{ik}\nabla_k\bar\phi'=0.
\label{eq3}
\end{equation}
Multiplying equation (\ref{eq3}) by $\delta r_i\mathrm{d}M_r$ and integrating from the center to the surface, it is not difficult after some simple derivations to obtain the amplitude growth rate, i.e. the normalized accumulated work
\begin{equation}
\eta=-\frac{2\pi\omega_i}{\omega_r}=-\frac{\pi}{2E}\int_{0}^{M_0}\mathrm{Im}\left\{-\frac{1}{\bar\rho}\left[\delta\left(\bar P+\bar\rho x^2\right)\frac{\delta\bar\rho^*}{\bar\rho}+\frac{1}{\bar\rho}\delta\left(\bar\rho\chi^{ik}\right)\nabla_k\delta r_i\right]\right\}\mathrm{d}M_r,
\label{eq4}
\end{equation}
where
\begin{equation}
E=\frac{1}{2}\omega^2\int_{0}^{M_0}\delta r^i\delta r_i\mathrm{d}M_r=\frac{1}{2}\omega^2\int_{0}^{M_0}\left[\delta r^2+l(l+1)\delta r_\mathrm{h}\right]\mathrm{d}M_r.
\label{eq5}
\end{equation}
From the energy equation, we obtain
\begin{equation}
\frac{\delta\bar P}{\bar P}=\Gamma_1\frac{\delta\bar\rho}{\bar\rho}+\frac{\Gamma_3-1}{P}\left\{\bar\rho x^2\left(\frac{\delta\bar\rho}{\bar\rho}-3\frac{\delta x}{x}\right)-\bar\rho\chi^{ik}\nabla_k\delta r_i-\frac{1}{i\omega}\nabla_k\left(F^k_\mathrm{r}+F^k_\mathrm{c}+F^k_\mathrm{t}\right)'\right\}.
\label{eq6}
\end{equation}
Substituting equation (\ref{eq6}), and equation (\ref{eq4}) becomes
\begin{eqnarray}
\eta=\frac{\pi}{2E}\int_0^{M_0}\mathrm{Im}\left\{(5-3\Gamma_3)\frac{\delta x^*}{x}\frac{\delta\bar\rho}{\bar\rho}+\delta\chi^{ik}\nabla_k\delta r^*_i+(\Gamma_3-2)\chi^{ik}\nabla_k\delta r_i\frac{\delta\bar\rho^*}{\bar\rho}\right\}\mathrm{d}M_r \nonumber\\
-\frac{\pi}{2E}\int_{0}^{M_0}\mathrm{Re}\left\{\frac{\Gamma_3-1}{4\pi r^3\bar\rho\omega}\nabla_k\left(F^k_\mathrm{r}+F^k_\mathrm{c}+F^k_\mathrm{t}\right)'\frac{\delta\bar\rho^*}{\bar\rho}\right\}\mathrm{d}M_r,
\label{eq7}
\end{eqnarray}
where $F_\mathrm{r}$, $F_\mathrm{c}$, and $F_\mathrm{t}$ are the radiative flux, convective energy flux, and turbulent kinetic energy flux, respectively. Generally speaking, $|F_\mathrm{t}| \ll |F_\mathrm{c}|$, therefore the influence of turbulent kinetic energy flux on pulsation stability is negligible, compared with thermal convection $F_\mathrm{c}$. In stellar pulsation, thermal convection always lags a little behind the change of density. Therefore thermal convection (the thermodynamic coupling between convection and oscillations) is generally a damping mechanism against oscillations. This gives rise to the red edge of the instability strip of Cepheids and Cepheid-like variables.

In equation (\ref{eq7}), the first term in the first curly brackets is the contribution of the isotropic component ($\rho x^2$) of turbulent Reynolds stress. It can be proved that $(5-3\Gamma_3)>0$, and turbulent pressure always lags a little behind the change of density. Therefore turbulent pressure is in general an excitation of oscillations. The second and third terms are the contribution of the anisotropic component ($\rho\chi^{ik}$) of turbulent Reynolds stress, i.e. the contribution of turbulent viscosity. Turbulent viscosity transforms regular pulsational kinetic energy into irregular turbulent kinetic energy. This process happens at low wavenumbers of turbulence (large-scale eddies). The turbulent kinetic energy of large-scale eddies transfers gradually to high-wavenumber eddies as a result of turbulence cascade, and is eventually transformed by molecular viscosity into thermal energy. Therefore turbulent anisotropy, i.e turbulent viscosity, is always a damping mechanism of oscillations.

We studied in detail the influence on stellar pulsation stability of the coupling between convection and oscillations \citep{XDZW2016}. As pointed out, the isotropic component of Reynolds stress, i.e. turbulent pressure, is always an excitation mechanism of stellar oscillations, while turbulent viscosity is always a damping mechanism. Turbulent viscosity is mostly included in the anisotropic component of the turbulent velocity correlation $\chi^{ij}$. When turbulent anisotropy is neglected, turbulent viscosity is disregarded at the same time. Therefore the red edge of the instability strip cannot be modelled correctly in the isotropic convection approximation in term of our theory.

We now turn to study quantitatively how stellar pulsation stability depends on the anisotropic parameter $c_3$ of turbulent convection. As we pointed out in Paper I, a quasi-anisotropic convection model can be constructed using non-local envelope models in isotropic convection approximation together with the so-called quasi-anisotropic approximation (equation (48) in Paper I). It approximates very well to the completely anisotropic convection model in terms of not only the $T-P$ structure, but also the turbulent velocity and temperature fields, as well as the properties of non-adiabatic oscillations. Therefore starting from non-local envelope models in isotropic convection approximation, we calculated six series of quasi-anisotropic convection models with different values of the convective parameter $c_3$ adopting the quasi-anisotropic approximation. The model parameters are given in Table \ref{tab1}, where the first column is the series number, the second column lists the values of the convective parameter $c_3$ used for each series, and the third column shows the ratio of squared radial velocity to squared horizontal velocity $\overline{u'^2_\mathrm{r}}/\overline{u'^2_\mathrm{h}}$. As $c_3$ increases, $\overline{u'^2_\mathrm{r}}/\overline{u'^2_\mathrm{h}}$ decreases monotonically, and eventually converges to $1/2$ in isotropic convection in convectively unstable zone. Thus isotropic convection can be regarded as the limit as $c_3 \to \infty$. In Paper I we discussed calibrations of convective parameters, and reached the conclusion that $c_3 \approx 3$ is a good choice. This means $\overline{u'^2_\mathrm{r}}/\overline{u'^2_\mathrm{h}}=1$ in convectively unstable zone, which corresponds to the linearly most unstable mode of \citet{Unno1967}. It is worth noting that our theory is a nonlocal convection theory, and $\overline{u'^2_\mathrm{r}}/\overline{u'^2_\mathrm{h}}$ is a variable instead of a constant. The radial component of the turbulent velocity dominates in the convectively unstable zone, while in the overshooting zone, the turbulent velocity is primarily horizontal \citep{DXC2006}.

\begin{figure}
\centering
\includegraphics[width=0.8\textwidth]{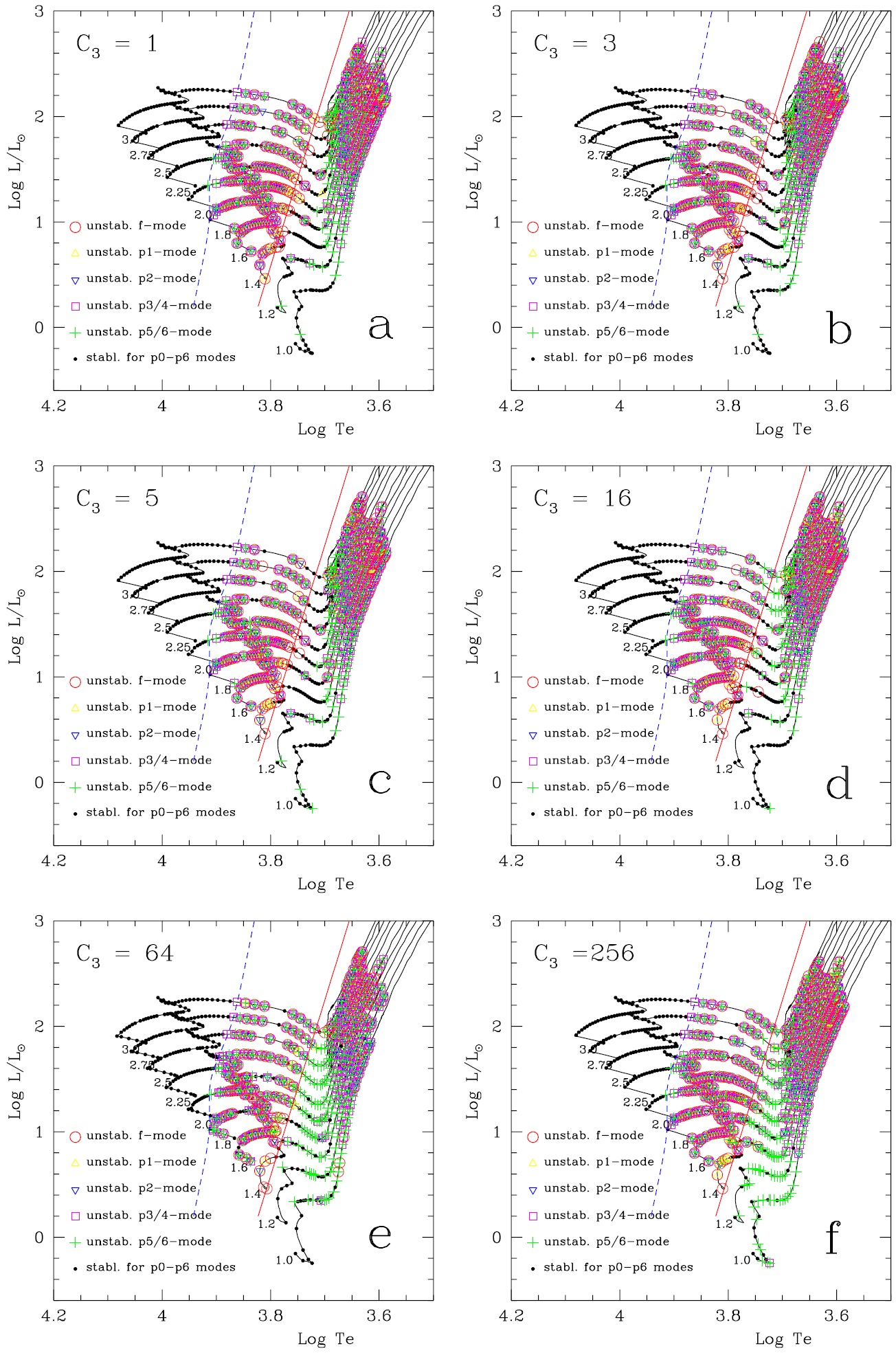}
\caption{Stable (black dots) and unstable (colored open symbols) low-order p modes of the six series of evolutionary models in the H-R diagram. The dashed and solid lines are respectively the theoretical blue and red edges of the $\delta$ Scuti instability strip. Panels a--f show series a--f of the evolutionary modes, respectively. The values of $c_3$ are given in Table \ref{tab1} and labeled in each panel.}
\label{fig2}
\end{figure}

\begin{figure}
\centering
\includegraphics[width=\textwidth]{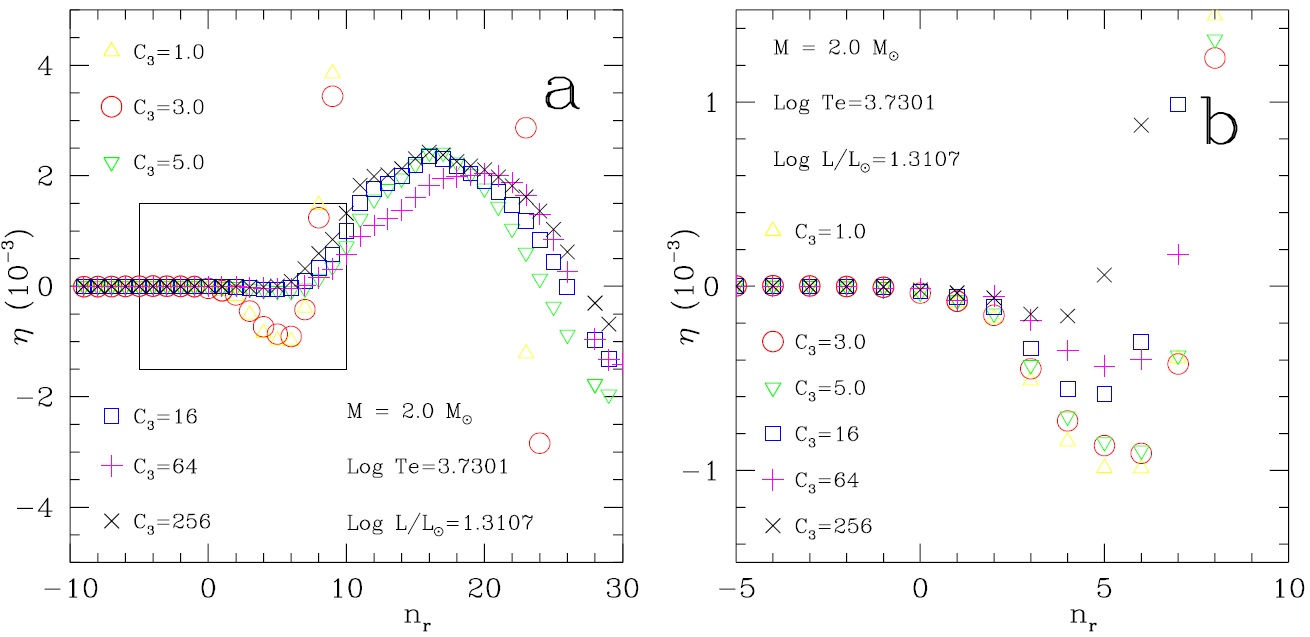}
\caption{Pulsation amplitude growth rate $\eta$ as a function of the radial order $n_\mathrm{r}$ with changing $c_3$ for a $2.0M_\odot$ low-temperature red star outside the red edge of the $\delta$ Scuti instability strip. Panel b is an enlarged view of the boxed area in Panel a.}
\label{fig3}
\end{figure}

Following the numerical scheme in Paper I, we calculated radial (p0--p39) and $l=2$ non-radial (g9--p29) oscillations for the six series of quasi-anisotropic convection models listed in Table \ref{tab1}. Figures \ref{fig2}a--f show the distribution of stable (black dots) and unstable (colored open symbols) f and low-order p modes for the six series of evolutionary models in the H-R diagram. As the convective parameter $c_3$ increases from 1 to 256, turbulent convection changes from highly anisotropic in Figure \ref{fig2}a to highly isotropic in Figure \ref{fig2}f, but pulsation stability of low-order ($n_\mathrm{r} < 5$) p modes shows no clear difference. Only the p5--p6 modes of low-temperature red stars outside the red edge of the $\delta$ Scuti instability strip become unstable, which is because the convection coupling has overtaken radiation $\kappa$ mechanism, and becomes the main excitation and damping mechanism of oscillations in these low-temperature red stars. The change of pulsation stability of low-order p modes with the convective parameter $c_3$ in Figures \ref{fig2}a--f reflects the influence of turbulent viscosity on stellar pulsation stability. Turbulent viscosity is caused by shear motion of fluid, and is mainly included in the anisotropic component. As $c_3$ increases, turbulent anisotropy decreases, therefore turbulent viscosity decreases, and stellar oscillations tend to become unstable. Figure \ref{fig3}a shows the pulsation amplitude growth rate $\eta$ of a low-temperature red star with $M=2.0M_\odot$ outside the red edge of the $\delta$ Scuti instability strip as a function of the radial order $n_\mathrm{r}$ with changing $c_3$. Figure \ref{fig3}b is an enlarged view of the boxed area in Figure \ref{fig3}a. It can be seen in Figure \ref{fig3} that pulsation stability of g and p modes with $n_\mathrm{r}<5$ hardly relies on the convective parameter $c_3$. However, toward higher radial order, p modes become unstable as $c_3$ increases. This perfectly explains the change of pulsation stability of low-order p modes with anisotropic parameter $c_3$ as shown in Figures \ref{fig2}a--f.

\begin{table}[h]
\begin{minipage}[]{100mm}
\caption{Model parameters\label{tab1}}
\setlength{\tabcolsep}{10pt}
\centering
\begin{tabular}{ccc}
\hline\noalign{\smallskip}
Series & $c_3$ & $\overline{u'^2_\mathrm{r}}/\overline{u'^2_\mathrm{h}}$\\
\hline\noalign{\smallskip}
a & 1.0 & 2.00\\
b & 3.0 & 1.00\\
c & 5.0 & 0.80\\
d & 16  & 0.59\\
e & 64  & 0.52\\
f & 256 & 0.51\\
\noalign{\smallskip}\hline
\end{tabular}
\end{minipage}
\end{table}

\begin{figure}
\centering
\includegraphics[width=0.9\textwidth]{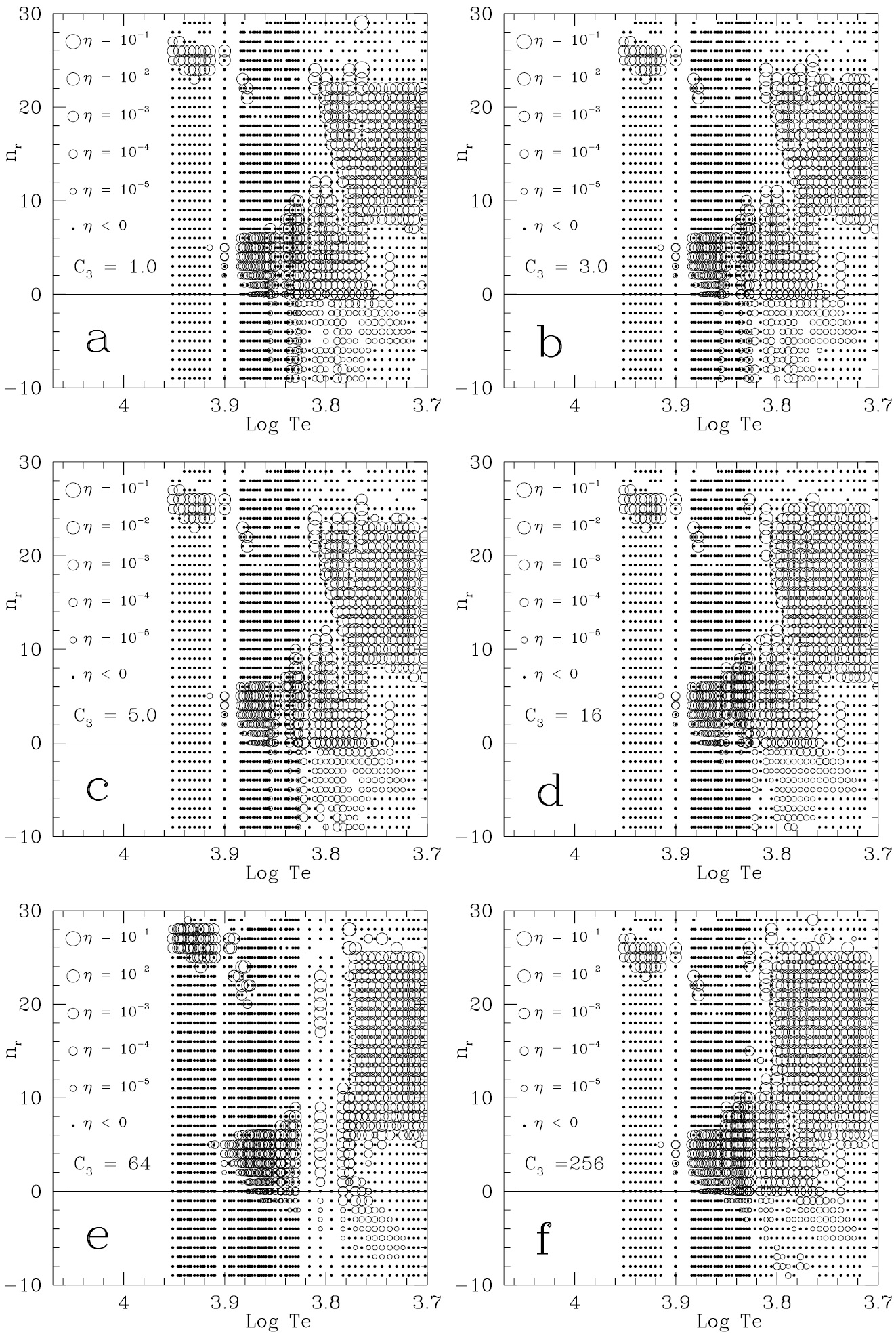}
\caption{Stable (black dots) and unstable (open circles) modes in the $n_\mathrm{r}-\log T_\mathrm{e}$ plane for a $2.0M_\odot$ star from the main sequence to the beginning of the RGB. The sizes of the open circles are proportional to $\log \eta$. Panels a--f show series a--f of the evolutionary modes, respectively.}
\label{fig4}
\end{figure}

\begin{figure}
\centering
\includegraphics[width=0.9\textwidth]{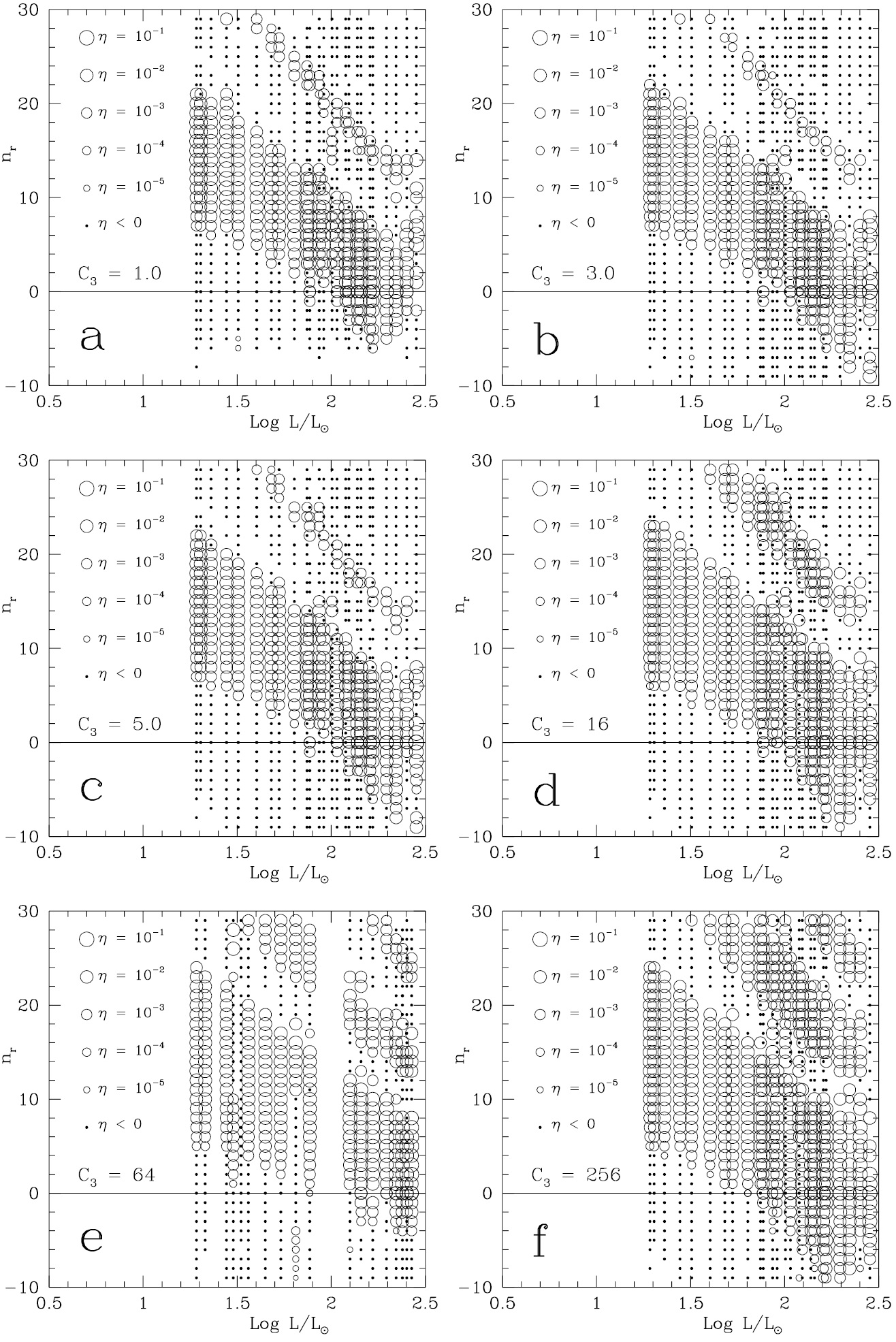}
\caption{Stable (black dots) and unstable (open circles) modes in the $n_\mathrm{r}-\log L/L_\odot$ plane for a $2.0M_\odot$ star along the RGB ($\log T_\mathrm{e}<3.70$). Panels a--f show series a--f of the evolutionary modes, respectively.}
\label{fig5}
\end{figure}

Figure \ref{fig3}a shows that, there is a bump in the pulsation amplitude growth rate $\eta$ as a function of the radial order $n_\mathrm{r}$. $\eta$ peaks at $n \sim 17$, and all $n=8$--$27$ modes are unstable. These are the so-called solar-like oscillations of stars \citep{XD2013}. The classical $\delta$ Scuti instability strip is defined in terms of low-order p modes. For high-order p modes, the $\delta$ Scuti instability strip and solar-like oscillations are connected to each other in the H-R diagram. Figures \ref{fig4}a--f show stable (black dots) and unstable (open circles) modes in the $n_\mathrm{r}-\log T_\mathrm{e}$ plane for $M=2.0M_\odot$ evolutionary models from main sequence to the beginning of the red-giant branch (RGB) of the six series with different convective parameter $c_3$, respectively. The sizes of the open circles are proportional to $\log \eta$. From Figures \ref{fig4}a--f, we can see:
\begin{enumerate}

\item There is a clear red edge of the $\delta$ Scuti-$\gamma$ Doradus instability strip for all g, f, and p modes with $n_\mathrm{r}<5$, and it hardly changes with the anisotropic parameter $c_3$.

\item There exists no red edge of the $\delta$ Scuti instability strip for p modes with $n_\mathrm{r}>8$. The instability strip and solar-like oscillations of low-temperature red stars joint together.

\item Figures \ref{fig4}a--f are very similar to each other. Pulsation stability does not show any drastic change
for models with different values of $c_3$, although there is a similar trend as in Figure \ref{fig3}. The areas of unstable high-order p modes in Figures \ref{fig4}e \& f are broader than those in Figures \ref{fig4}a--d.

\end{enumerate}

Figures \ref{fig4}a--f show only the change of pulsation stability with $\log T_\mathrm{e}$ and $c_3$ for a $2.0M_\odot$ star before entering the RGB phase, while Figures \ref{fig5}a--f show stable (black dots) and unstable (open circles) modes in the $n_\mathrm{r}-\log L/L_\odot$ plane for the six series of $2.0M_\odot$ evolutionary models along the RGB ($\log T_\mathrm{e}<3.70$). Figures \ref{fig5}a--f are very similar, only that the areas of unstable high-order p modes in Figures \ref{fig5}e--f are broader than those in Figures \ref{fig5}a--d. This means pulsation stability is also not sensitive to turbulent anisotropy for RGB stars.

\section{Conclusion and Discussion}
\label{sect:conclusions}
In this work we have studied in detail the dependence of pulsation stability of $\delta$ Scuti/$\gamma$ Doradus stars on the anisotropic parameter $c_3$ of turbulent convection. The results show that:
\begin{enumerate}

\item Turbulent anisotropy has virtually no influence on pulsation stability of g modes and low-order ($n_\mathrm{r}<5$) p modes. In a wide range of the anisotropic parameter $c_3$ ($1 \lesssim c_3 \lesssim 16$), pulsation stability of g and low-order p modes hardly depends on $c_3$. Therefore we are confident to say that, calibration errors of $c_3$ have no substantial effect on pulsation stability of $\delta$ Scuti/$\gamma$ Doradus stars.

\item The effect of turbulent anisotropy on pulsation stability of high-order ($n_\mathrm{r} \gtrsim 5$) p modes is non-negligible. Most of high-order p modes of all low-temperature red stars become unstable if turbulent anisotropy is ignored, and the red edge of the $\delta$ Scuti instability strip cannot be modelled theoretically.

\end{enumerate}

Our non-local and anisotropic model and 3D hydrodynamical simulations show good agreement in the deep convection zone, but have notable difference near the boundary and in the overshooting zone. However, anisotropy has no substantial influence on pulsation stability of g modes and low-order p modes, and the overshooting zone contributes little to mode inertia. Therefore, at least for g modes and low-order p modes, the uncertainty of turbulent anisotropy $\overline{u'^2_\mathrm{h}}/\overline{u'^2_\mathrm{r}}$ in the overshooting zone in static models has no substantial influence on pulsation stability.

\section*{Acknowledgements}
This work is supported by National Natural Science Foundation of China (NSFC) through grants 11403039, 11473037, and 11373069.

\end{document}